\documentclass{jcgt}

\setciteauthor{Russell A. Brown}
\setcitetitle{Comparative Performance of the AVL Tree and Three Variants of the Red-Black Tree}
\accepted{submitted}{accepted}{published}{Editor Name}{vol}{issue}{1}{1}{year}
\seturl{http://jcgt.org/published/vol/issue/num/}

\usepackage[usenames,dvipsnames]{xcolor}
\usepackage{tcolorbox}
\usepackage{tabularx}
\usepackage{array}
\usepackage{tabto}
\usepackage{colortbl}
\tcbuselibrary{skins}

\newcolumntype{Y}{>{\raggedleft\arraybackslash}X}
\tcbset{tab1/.style={fonttitle=\bfseries\large,fontupper=\normalsize\sffamily,
colback=yellow!10!white,colframe=red!75!black,colbacktitle=Salmon!40!white,
coltitle=black,center title,freelance,frame code={
\foreach \n in {north east,north west,south east,south west}
{\path [fill=red!75!black] (interior.\n) circle (3mm); };},}}

\tcbset{tab2/.style={enhanced,fonttitle=\bfseries,fontupper=\normalsize\sffamily,
colback=white!10!white,colframe=black!50!black,colbacktitle=Salmon!40!white,
coltitle=black,center title}}

\begin{document}

\title{Comparative Performance of the AVL Tree and \linebreak Three Variants of the Red-Black Tree}

\author
       {Russell A. Brown}

\maketitle

\begin{abstract}
\small

This article compares the performance of the AVL tree to the performance of the bottom-up, top-down, and left-leaning red-black trees. The bottom-up red-black tree is faster than the AVL tree for insertion and deletion of randomly ordered keys. The AVL tree is faster than the bottom-up red-black tree for insertion but slower for deletion of consecutively ordered keys. The top-down red-black tree is faster than the bottom-up red-black tree for insertion but slower for deletion of randomly ordered keys, and slower for insertion and deletion of consecutively ordered keys. The left-leaning red-black tree is slower than the three other trees for insertion and deletion of randomly and consecutively ordered keys. An alternative deletion algorithm, which reduces the number of rebalancing operations required by deletion, is analyzed.

\end{abstract}

%-------------------------------------------------------------------------
% Remove the following asterisk and all asterisks that follow \section to make the sections visible in the page header.
\section{AVL Tree}
\label{sec:AVL}

The AVL tree is a special form of binary search tree (BST) that guarantees $ O\left( \log n \right ) $ insertion, deletion, and search. This guarantee is obtained at the cost of rebalancing the AVL tree, potentially after each insertion \cite{Adelson} or deletion \cite{Foster}.

\subsection{AVL Tree Balance and Rotation}
\label{sec:Balance}

At each node of an AVL tree, the node's left and right subtrees are allowed to differ in height, which is defined as the maximum path length to the bottom of the tree, by at most one node. This difference is expressed at each node via a \lstinline{balance} field whose allowed value of -1, 0, or +1 represents the height of the right subtree minus the height of the left subtree. If, after insertion or deletion of a node, \lstinline{balance} equals -2 or +2 at any higher node in the tree, the subtree rooted at that higher node is rebalanced by cyclically exchanging that higher node with 1 or 2 of its subordinate nodes to establish a new root node for the subtree and reset its \lstinline{balance} field to zero \cite{Tsakalidis}.

One of four distinct rotation operations cyclically exchanges the nodes. The four rotation operations are the left-left (LL) and right-right (RR) single rotations that exchange two nodes, and the left-right (LR) and right-left (RL) double rotations that exchange three nodes \cite{Wirth} \cite{Drozdek} \cite{Weiss}.

\subsection{AVL Tree Deletion}
\label{sec:AVLDeletion}

A customary algorithm to delete an AVL node from the AVL tree first treats the tree as if it were a general BST to delete the node, and then rebalances the tree if necessary \cite{Wirth} \cite{Drozdek} \cite{Weiss}. Deletion of an AVL node that is treated as if it were a BST node occurs as follows. If the node has no children, it is removed from the tree. If the node has only one child, the node is replaced by its child. If the node has two children, the node is replaced either by its in-order predecessor (i.e., the rightmost node of the left subtree) or by its in-order successor (i.e., the leftmost node of the right subtree) \cite{Hibbard} \cite{Knuth}.

The general BST contains no information to guide a choice between the rightmost node of the left subtree and the leftmost node of the right subtree. Hence, for the BST, either node is a viable replacement for the deleted node. In contrast, each node of the AVL tree contains a \lstinline{balance} field that can guide selection of a preferred replacement node when the deleted AVL node has two children \cite{Foster}.

If \lstinline{balance} equals -1, the height of the left subtree exceeds the height of the right subtree, so the deleted node is replaced by the rightmost node of the left subtree. But if \lstinline{balance} equals +1, the height of the right subtree exceeds the height of the left subtree, so the deleted node is replaced by the leftmost node of the right subtree. And if \lstinline{balance} equals 0, the subtrees have equal heights, so the deleted node may be replaced either by the rightmost node of the left subtree or by the leftmost node of the right subtree. This alternative node replacement algorithm often avoids rebalancing after deletion because it replaces the deleted AVL node with a preferred node from the taller of the two subtrees, which may shorten the taller subtree, thereby avoiding the need to rebalance.

\begin{figure}[h]
\centering
\centerline{\includegraphics*[trim = {0.94in, 0.96in, 0.94in, 0.96in}, clip, width=\textwidth]{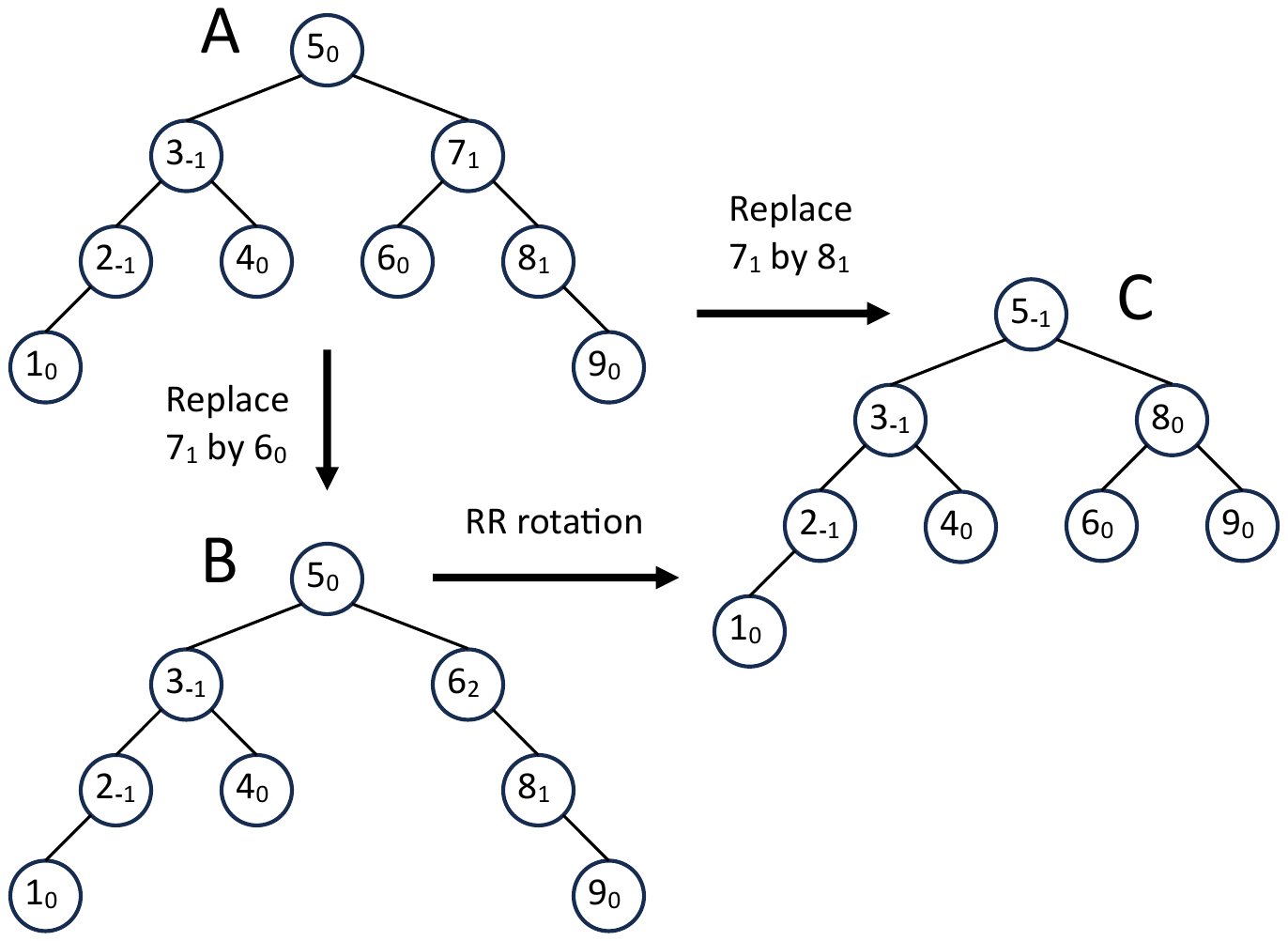}}
\caption{Replacement of node $7_1$ either by preferred node $8_1$ or by unpreferred node $6_0$}
\label{fig:rotation}
\end{figure}

Figure \ref{fig:rotation} illustrates the alternative node replacement algorithm for the AVL tree depicted in Figure \ref{fig:rotation}A. In this AVL tree, which contains keys 1 through 9, each AVL node's label specifies the node's \lstinline{key} and \lstinline{balance} fields. For example, \lstinline{key}$\;=5$ and \lstinline{balance}$\;=0$ at node $5_0$, and \lstinline{key}$\;=7$ and \lstinline{balance}$\;=+1$ at node $7_1$. Figures \ref{fig:rotation}B and \ref{fig:rotation}C depict trees from which node $7_1$ was removed and replaced by nodes $6_0$ and $8_1$ respectively. Figure \ref{fig:rotation}A shows that at node $7_1$, the right subtree is taller than the left subtree, as indicated also by \lstinline{balance}$\;=+1$ at node $7_1$, so the preferred replacement node is the leftmost node of the right subtree (i.e., node $8_1$). Figure \ref{fig:rotation}C shows that replacing node $7_1$ by preferred node $8_1$ preserves the balance of the tree. In contrast, Figure \ref{fig:rotation}B reveals that replacing node $7_1$ by the rightmost node of the left subtree (i.e., unpreferred node $6_0$) produces an unbalanced subtree, as indicated by \lstinline{balance}$\;=+2$ at node $6_2$, so that subtree must be rebalanced via an RR rotation that cyclically exchanges nodes $6_2$ and $8_1$ to produce the tree depicted in Figure \ref{fig:rotation}C.

Although the example of Figure \ref{fig:rotation}A depicts short subtrees rooted at node $7_1$, the choice of a preferred replacement node via \lstinline{balance} for subtrees of any height may reduce the number of rotations required by deletion of a node that has two children.

\section{Red-Black Tree}
\label{sec:RedBlack}

The red-black tree is another special form of BST that guarantees $ O\left( \log n \right ) $ insertion, deletion, and search, at the cost of rebalancing via rotations, i.e., cyclic node exchanges, that are identical to those of the AVL tree \cite{Guibas}. The red-black tree is an improvement to the symmetric binary B-tree (SBB tree) \cite{Bayer} because each node of an SBB tree requires one bit for each of its two child pointers to represent that pointer's color (red or black), whereas each node of a red-black tree requires only one bit to specify that node's color \cite{Wirth85}. In contrast to the AVL tree that maintains a \lstinline{balance} field at each node to specify the balance of the subtree rooted at that node, the red-black tree maintains a \lstinline{color} field at each node that does not specify the balance of the subtree. Instead, the red-black tree promotes balance by enforcing the following constraints: (1) the root node is black; (2) if a node is red, both of its children must be black; and (3) every path from a given node to the bottom of the tree must contain the same number of black nodes \cite{Weiss03}.

Given the above color constraints, the shortest path from the root to the bottom of the red-black tree comprises only black nodes, whereas the longest path comprises alternating red and black nodes. Hence, because every path from the root to the bottom of the tree must contain the same number of black nodes, the longest path is twice the length of the shortest path \cite{Drozdek}.

The above color constraints further imply that, if every path from the root to the bottom of the red-black tree contains $b$ black nodes, a tree that contains only black nodes comprises $n=2^b-1$ nodes, and the height $h$ of this tree along any path to the bottom of the tree is $h = \log_2 \left( n+1 \right)$ \cite{Weiss03}. Moreover, for this tree that contains only black nodes, all paths have equal length and that length is the shortest path to the bottom of the tree. Adding the maximum allowed number of red nodes along any path of this tree, thus creating a path of alternating red and black nodes, achieves a longest path whose length is double the length of the shortest path, i.e., $h = 2\log_2 \left( n+1 \right)$. So, for either the shortest or longest path, descent from the root to the bottom of the tree (such as for insertion, deletion, or search) is $ O\left[ \log_2 \left( n+1 \right ) \right] $.

Three variants of the red-black tree are discussed subsequently.

\subsection{Bottom-up and Top-down Red-Black Trees}
\label{sec:BottomUp}

Rebalancing a red-black tree may be performed in either a bottom-up or a top-down manner \cite{Guibas}. Bottom-up rebalancing is similar to rebalancing an AVL tree, wherein the tree is searched downward for a point of insertion or deletion of a node, and then the tree is rebalanced by rotation, if necessary, at each node encountered when backtracking upward from the point of insertion or deletion towards the root of the tree \cite{Weiss03}.

Top-down rebalancing occurs as the tree is searched downward for a point of insertion or deletion, thereby avoiding the need to rebalance along an upward path after insertion or deletion \cite{Weiss03}, and also avoiding the need for a pointer from each node to its parent node to enable upward traversal of the tree \cite{Tarjan}. Top-down insertion is simple but top-down deletion is complicated \cite{Weiss03}.

\subsection{Left-leaning Red-Black Tree}
\label{sec:LeftLeaning}

The left-leaning red-black tree \cite{Sedgewick} is a modification of the AA tree, which is an asymmetric variant of the binary B-tree \cite{Andersson}. The left-leaning red-black tree is an improvement to the AA tree because each node of the AA tree requires a \lstinline{height} field of some number of bits \cite{Weiss03}, whereas each node of the left-leaning red-black tree requires only one bit to represent that node's \lstinline{color} field \cite{Sedgewick}. Implementation of a left-leaning red-black tree is simpler than implementation of a bottom-up or top-down red-black tree because the left-leaning red-black tree's asymmetry introduces fewer distinct cases that require analysis and rebalancing. However, whereas insertion into the left-leaning red-black tree requires only top-down rebalancing, deletion from the left-leaning red-black tree requires both top-down and bottom-up rebalancing. No pointer from each node to its parent node is required for bottom-up rebalancing because the recursive deletion algorithm performs bottom-up rebalancing as the recursion unwinds.

\subsection{Red-Black Tree Deletion}
\label{sec:RedBlackDeletion}

A customary algorithm to delete a red-black node from a red-black tree resembles the customary deletion algorithm for an AVL tree because the algorithm treats the red-black tree as if it were a general BST to delete the node \cite{Drozdek} \cite{Sedgewick11}. Given this resemblance, the alternative node replacement algorithm proposed for the AVL tree in Section \ref{sec:AVLDeletion} may be applied to deletion of a node that has two children from a red-black tree, and thereby permit selection of a preferred replacement node for that deleted two-child node. Although each node of a red-black tree does not contain a \lstinline{balance} field to guide selection of a preferred replacement node, each node of the tree could maintain a \lstinline{size} field to represent the number of nodes in the subtree rooted at that node \cite{Sedgewick11}. This \lstinline{size} field could substitute for the \lstinline{balance} field to guide selection of a preferred replacement node by selecting that node from the larger subtree.

Maintaining a \lstinline{size} field is simple for a bottom-up red-black tree, because that tree's insertion and deletion algorithms backtrack upward from the point of insertion or deletion to the root of the tree. At each node visited along the upward path, the \lstinline{size} field is computed by adding 1 to the sum of the \lstinline{size} fields of the node's two children. The alternative node replacement algorithm often avoids rebalancing after deletion of a two-child node from a bottom-up red-black tree, similar to the avoidance of rebalancing for the AVL tree.

It is possible to maintain a \lstinline{size} field for a left-leaning red-black tree \cite{Argento}. However, the node replacement algorithm is not viable for that tree because that tree's asymmetry precludes replacement of a two-child node by its in-order predecessor, but instead permits replacement of that node by only its in-order successor.

Maintaining a \lstinline{size} field might be possible for a top-down red-black tree using the same approach used for the left-leaning red-black tree \cite{Argento}. However, the customary deletion algorithm for the top-down red-black tree is substantially slower than the customary deletion algorithm for the bottom-up red-black tree (see Figure \ref{fig:RandomDeletionTimes} below). Hence, even if the alternative node deletion algorithm improved deletion performance for the top-down red-black tree, its performance would likely still be substantially slower than the deletion performance for the bottom-up red-black tree.

\newpage

\section{Benchmarks}
\label{sec:Benchmarks}

\subsection{Benchmark Methodology}
\label{sec:Methodology}

To compare the performance of the AVL and red-black trees, benchmarks were executed on a Hewlett-Packard Pro Mini 400 G9 with 2x16GB DDR5-5600 RAM and a 14th-generation Intel Raptor Lake CPU (i7 14700T with 8 performance cores, 5.2GHz performance core maximum frequency, 89.6GB/s maximum memory bandwidth, 80KB L1 and 2MB per-core L1 and L2 caches, and a 33MB L3 shared cache).
 
A benchmark for each type of tree was implemented in C++, compiled via Gnu g++ 13.2.0 with the \lstinline{-O3} and \lstinline{-Winline} options, and executed under Ubuntu 24.04.1 LTS as a single thread on a single performance core specified via the Ubuntu \lstinline{taskset} command. No other applications were executed concurrently.

Benchmarks for each type of tree were executed for a set of trees wherein each node stored a 32-bit integer key. Each tree in the set comprised a number of nodes $n$ equal to an integer power of two in the range $ \left [ 2^{16} , 2^{20} \right] $. To ensure reliable statistics (defined as the standard deviation of an execution time less than 3 percent of the execution time), each execution performed a number of iterations $ 2_i^n $ where $i$ specifies the exponent of the number of iterations and $n$ specifies the exponent of the number of nodes. Thus, the trees were benchmarked via $ \left\{ 2_{14}^{16} , 2_{13}^{17} , 2_{12}^{18} , 2_{11}^{19} , 2_{10}^{20} \right\}$ executions.

For each iteration of one series of benchmarks, consecutive integers were randomly shuffled prior to insertion and then randomly shuffled again prior to deletion via the \lstinline{std::mt19937_64} Mersenne Twister pseudo-random number generator \cite{Matsumoto}. Each benchmark initialized \lstinline{std::mt19937_64} to \lstinline{std::mt19937_64::default_seed} so that all benchmarks randomly shuffled the integers in an identical sequence. For each iteration of a second series of benchmarks, the consecutive integers were inserted and deleted in increasing sorted order.

For the benchmarks, the node size for each type of tree was 32 bytes that comprised: (1) one 8-byte word for each of the parent and child pointers, i.e., 24 bytes; and (2) one 8-byte word that contained a 4-byte integer key and a 1-byte \lstinline{balance} or \lstinline{color} field. Although only the bottom-up red-black tree requires a parent pointer for insertion or deletion, a parent pointer facilitates iteration for a BST \cite{Pfaff}. Hence, the node for each type of tree included a parent pointer. Also, to support alternative deletion for the bottom-up red-black tree, each node of that tree included one 8-byte word for the \lstinline{size} field when alternative deletion was specified.

For each iteration, the execution times for insertion and deletion were measured via the \lstinline{std::chrono::steady_clock::now()} function. For each iteration, the numbers of rotations were counted for insertion and deletion. Each single rotation was counted as one rotation. Each double rotation was counted as two rotations because a double rotation is composed of two single rotations.

\subsection{Rotation Counts for Random-order Insertion and Deletion}
\label{sec:rotationcounts}

\begin{figure}[h]
\centering
\centerline{\includegraphics*[trim = {0.97in, 3.47in, 1.36in, 1.51in}, clip, width=\textwidth]{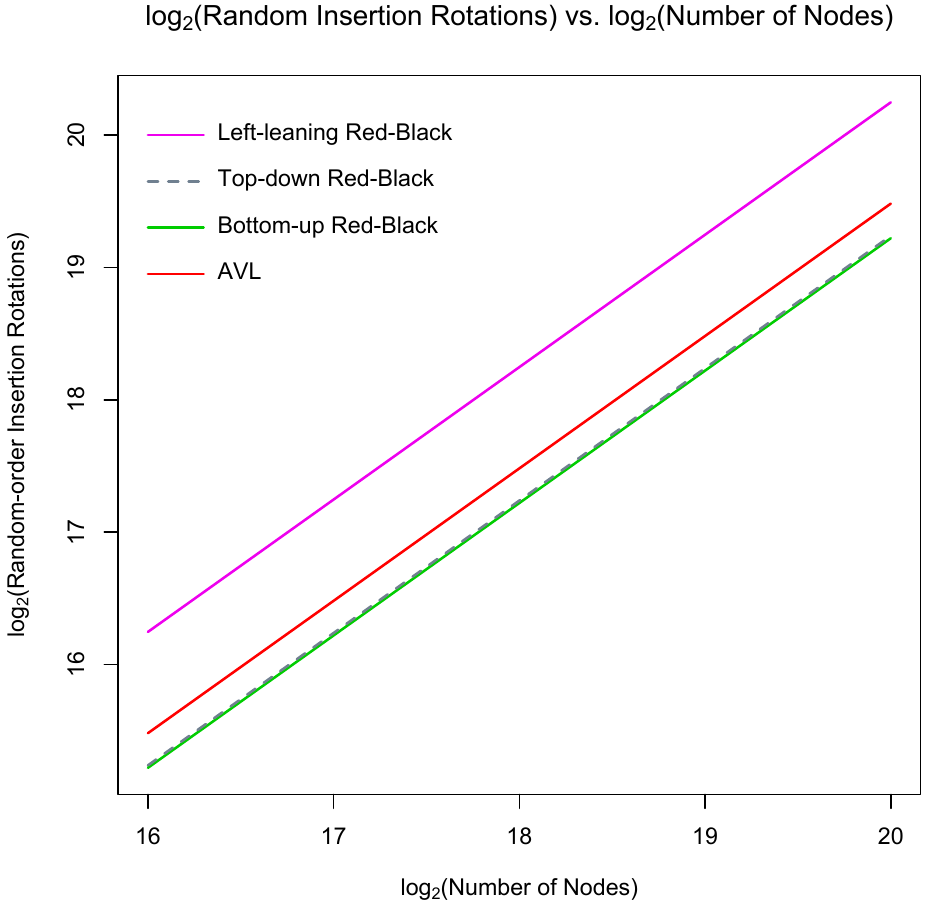}}
\caption{Number of rotations required by insertion versus the number of nodes}
\label{fig:insertionrotations}
\end{figure}

Figure \ref{fig:insertionrotations} shows log-log plots of the number of rotations required by random-order insertion into each tree in the set for each type of tree, versus the number of nodes in the tree. The $y$-axis displacement between plots in log space corresponds to multiplication in linear space, and reveals the relative numbers of rotations for the trees. The top-down red-black (gray), AVL (red), and left-leaning red-black (magenta) plots are displaced upward by 0.02, 0.26, and 1.03 $ \log_2 $ units respectively relative to the bottom-up red-black (green) plot, so those trees' insertion algorithms respectively perform $ 2^{0.02} = 1.01 $, $ 2^{0.26} = 1.20 $, and $ 2^{1.03} = 2.04 $ times more rotations than the bottom-up red-black tree. (The displacements are calculated from insertion rotation data for $ 2^{18} $ nodes instead of measured from Figure \ref{fig:insertionrotations}.)

\newpage

\begin{figure}[h]
\centering
\centerline{\includegraphics*[trim = {0.97in, 3.47in, 1.36in, 1.51in}, clip, width=\textwidth]{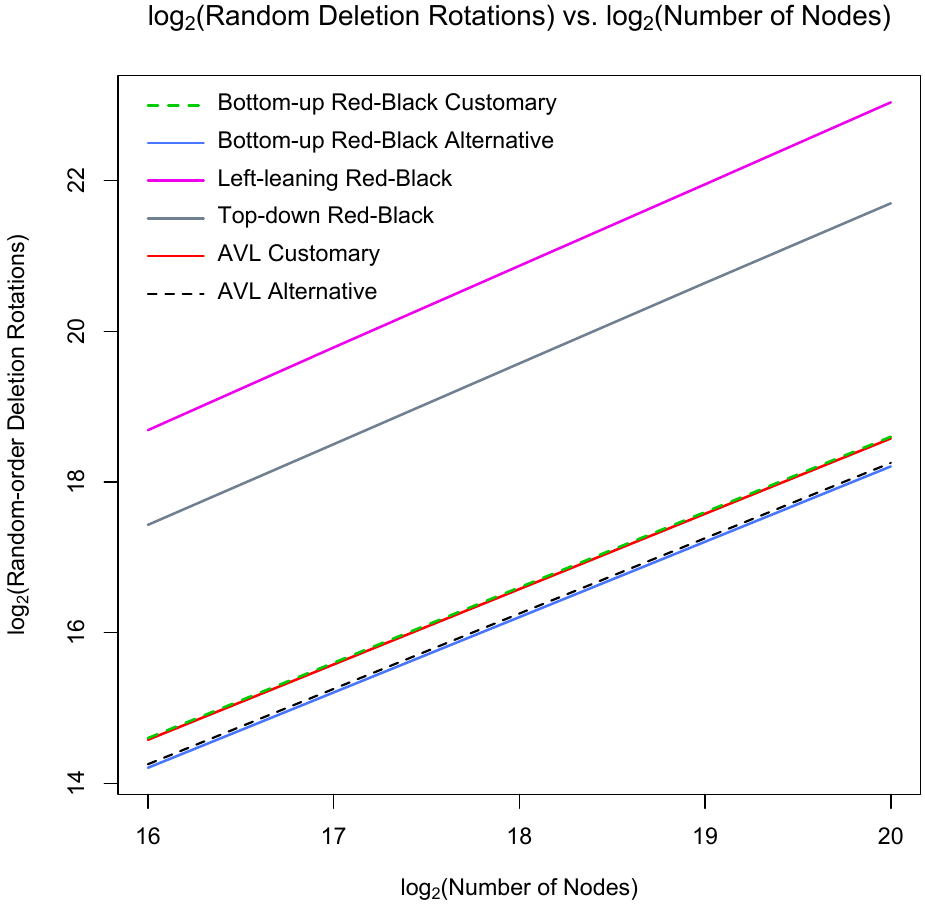}}
\caption{Number of rotations required by deletion versus the number of nodes}
\label{fig:deletionrotations}
\end{figure}

Figure \ref{fig:deletionrotations} shows log-log plots of the number of rotations required by random-order deletion from each tree in the set for each type of tree, versus the number of nodes in the tree. The $y$-axis displacement between plots in log space, calculated from deletion rotation data for $ 2^{18} $ nodes, reveals that the AVL alternative (black), AVL customary (red), bottom-up red-black customary (green), top-down red-black (gray), and left-leaning red-black (magenta) trees' deletion algorithms respectively perform $ 2^{0.05} = 1.03 $, $ 2^{0.37} = 1.29 $, $ 2^{0.39} = 1.31 $, $ 2^{3.37} = 10.3 $, and $ 2^{4.66} = 25.3 $ times more rotations than the bottom-up red-black tree's alternative deletion algorithm (blue).

The $y$-axis displacement between plots in log space, calculated from deletion rotation data for $ 2^{18} $ nodes, reveals that the AVL tree's customary deletion algorithm (red) performs $ 2^{0.32} = 1.25 $ times more rotations than its alternative deletion algorithm (black), and that the bottom-up red-black tree's customary deletion algorithm (green) performs $ 2^{0.39} = 1.31 $ times more rotations than its alternative deletion algorithm (blue). Hence, the AVL and bottom-up red-black trees' alternative deletion algorithms require respectively $ 1 - 1 / 1.25 = 0.20 $ and $ 1 - 1 / 1.31 = 0.24 $, or 20 and 24 percent, fewer rotations than the corresponding customary deletion algorithms.

\subsection{Execution Times for Random-order Insertion and Deletion}
\label{sec:RandomOrder}

\begin{figure}[h]
\centering
\centerline{\includegraphics*[trim = {0.97in, 3.47in, 1.36in, 1.51in}, clip, width=\textwidth]{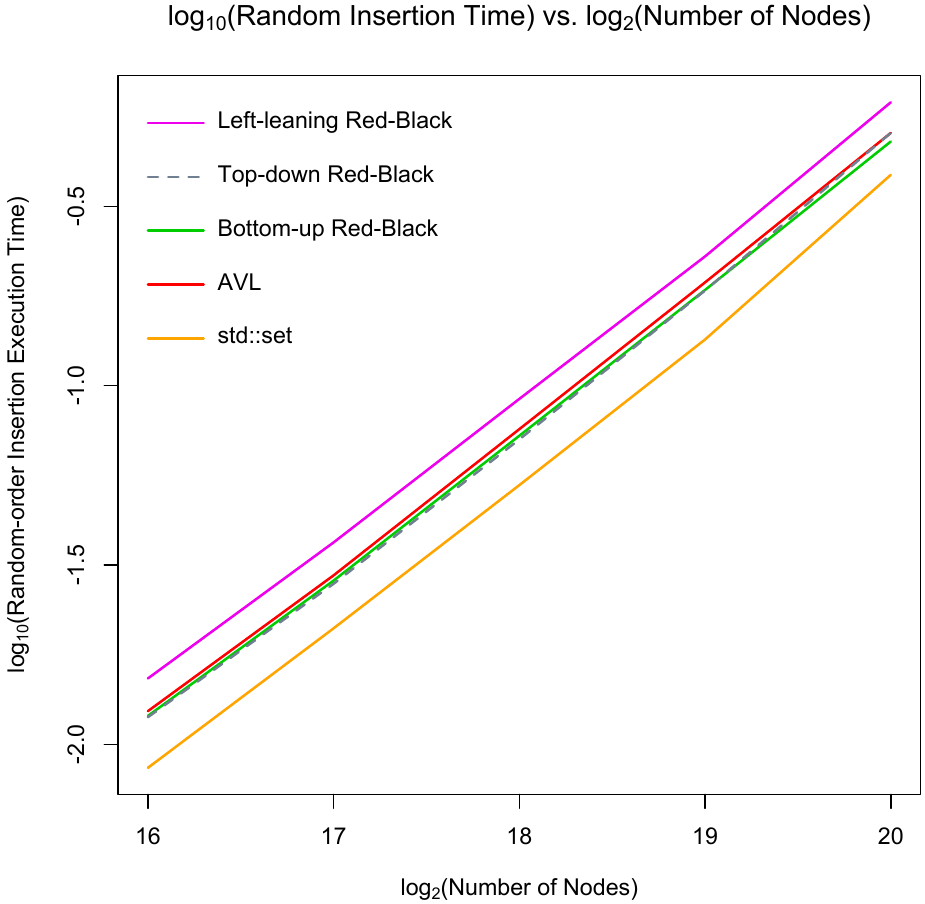}}
\caption{Random-order insertion execution times versus the number of nodes}
\label{fig:RandomInsertionTimes}
\end{figure}

Figure \ref{fig:RandomInsertionTimes} shows log-log plots of the Raptor Lake i7 CPU execution times for random-order insertion into each tree in the set for each type of tree and for \lstinline{std::set} (which served as a reference), versus the number of nodes. The $y$-axis displacement between plots in log space, calculated from insertion execution times for $ 2^{18} $ nodes, reveals that the top-down red-black tree's insertion algorithm (gray) is respectively $ 10^{0.01} = 1.02 $, $ 10^{0.03} = 1.07 $, and $ 10^{0.11} = 1.30 $ times faster than the bottom-up red-black (green), AVL (red), and left-leaning red-black (magenta) trees' insertion algorithms.

\newpage

\begin{figure}[h]
\centering
\centerline{\includegraphics*[trim = {0.97in, 3.47in, 1.36in, 1.51in}, clip, width=\textwidth]{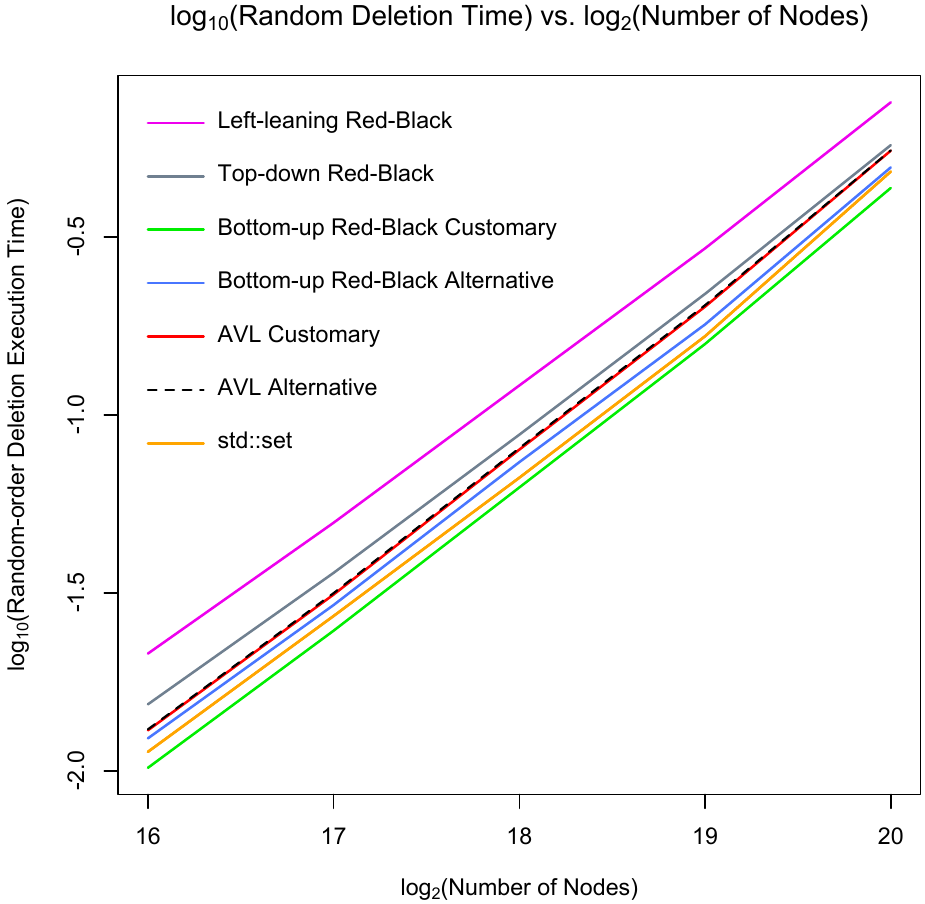}}
\caption{Random-order deletion execution times versus the number of nodes}
\label{fig:RandomDeletionTimes}
\end{figure}

Figure \ref{fig:RandomDeletionTimes} shows log-log plots of the Raptor Lake i7 CPU execution times for random-order deletion from each tree in the set for each type of tree and for \lstinline{std::set}, versus the number of nodes. The $y$-axis displacement between plots in log space, calculated from deletion execution times for $ 2^{18} $ nodes, reveals that the bottom-up red-black (green) tree's customary deletion algorithm is respectively $ 10^{0.07} = 1.18 $, $ 10^{0.106} = 1.28 $, $ 10^{0.11} = 1.29 $, $ 10^{0.15} = 1.41 $, and $ 10^{0.29} = 1.93 $ times faster than the bottom-up red-black alternative (blue), AVL customary (red), AVL alternative (black), top-down red-black (gray), and left-leaning red-black (magenta) trees' deletion algorithms. The displacement for $ 2^{18} $ nodes also reveals that the AVL tree's customary deletion algorithm is $ 10^{0.004} = 1.01 $ times faster than its alternative deletion algorithm.  The AVL and bottom-up red-black trees' alternative deletion algorithms are slower than their customary deletion algorithms, despite requiring fewer rotations; this result will be discussed in Section \ref{sec:AlternativeDeletion}.

Comparison of Figure \ref{fig:RandomInsertionTimes} to Figure \ref{fig:insertionrotations}, and comparison of Figure \ref{fig:RandomDeletionTimes} to Figure \ref{fig:deletionrotations}, reveal that, in general,  the insertion and deletion execution times are directly correlated to the number of rotations, as predicted previously \cite{Wirth85}. Exceptions to this direct correlation are the alternative deletion algorithms for the AVL and bottom-up red-black trees. Section \ref{sec:AlternativeDeletion} will discuss these exceptions.

The benchmark results for random-order insertion into and deletion from the AVL and red-black trees are summarized in Table \ref{table:RandomInsertionDeletionSpeed} that reports the relative speeds of the AVL and red-black trees for insertion and deletion of $ 2^{18} $ nodes. For insertion, the top-down red-black tree is fastest. For deletion, the bottom-up red-black tree's customary deletion algorithm is fastest. The relative speeds of the AVL and bottom-up red-black trees are consistent with the results of previously reported benchmarks \cite{Pfaff}.

% Create a narrower column type W to accommodate "Left-leaning" in the following table and
% then make a rightmost YYYY column to that table to suppress a rightmost vertical line
% that appears when the Z column type is substituted for the default X column type.
% In fact, a rightmost YY column suffices but additional Y characters shrink the space
% between the rightmost character of the table and the surrounding tcolorbox.
\newcolumntype{W}{>{\hsize=0.60\hsize}X}

\begin{table}[htb]

\begin{tcolorbox}[tab2,tabularx={W||Y|Y|YYYY}]
Algorithm & Insertion & Customary Deletion & Alternative Deletion  \\\hline\hline
AVL & 1.07 & 1.28 &1.29 \\\hline
Bottom-up \newline Red-Black & 1.02 & 1.00 & 1.18 \\\hline
Top-down \newline Red-Black & 1.00 & 1.41  \\\hline
Left-leaning \newline Red-Black & 1.30 & 1.93  \\\hline
\end{tcolorbox}

\caption{\label{table:RandomInsertionDeletionSpeed}
Relative speeds for random-order insertion and deletion of $ 2^{18} $ nodes}

\end{table}

\subsection{Execution Times for Sorted-order Insertion and Deletion}
\label{sec:SortedOrder}

Previously reported benchmarks report that ``if insertions often occur in sorted order, AVL trees excel when later accesses tend to be random" \cite{Pfaff}. To investigate this assertion, a second series of benchmarks was performed wherein the keys were inserted into and deleted from the AVL and red-black trees in consecutive order (i.e., increasing sorted order).

\newpage

\begin{figure}[h]
\centering
\centerline{\includegraphics*[trim = {0.97in, 3.47in, 1.36in, 1.51in}, clip, width=\textwidth]{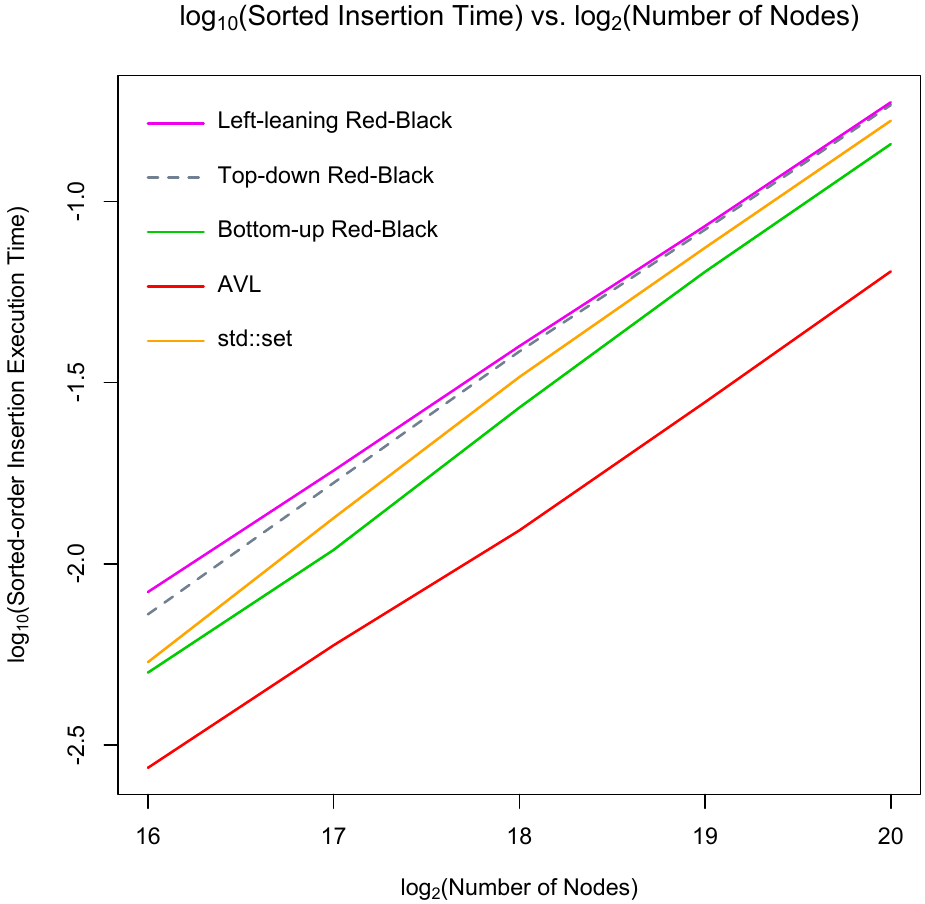}}
\caption{Sorted-order insertion execution times versus the number of nodes}
\label{fig:SortedInsertionTimes}
\end{figure}

Figure \ref{fig:SortedInsertionTimes} shows log-log plots of the Raptor Lake i7 CPU execution times for sorted-order insertion into each tree in the set for each type of tree and for \lstinline{std::set}, versus the number of nodes. The $y$-axis displacement between plots in log space, calculated from insertion execution times for $ 2^{18} $ nodes, reveals that the AVL tree's insertion algorithm (red) is respectively $ 10^{0.34} = 2.18 $, $ 10^{0.49} = 3.12 $, and $ 10^{0.51} = 3.22 $ times faster than the bottom-up red-black (green), top-down red-black (gray), and left-leaning red-black (magenta) trees' insertion algorithms.

Hence, the AVL tree outperforms the bottom-up, top-down, and left-leaning red-black trees for sorted-order insertion.

\newpage

\begin{figure}[h]
\centering
\centerline{\includegraphics*[trim = {0.97in, 3.47in, 1.36in, 1.51in}, clip, width=\textwidth]{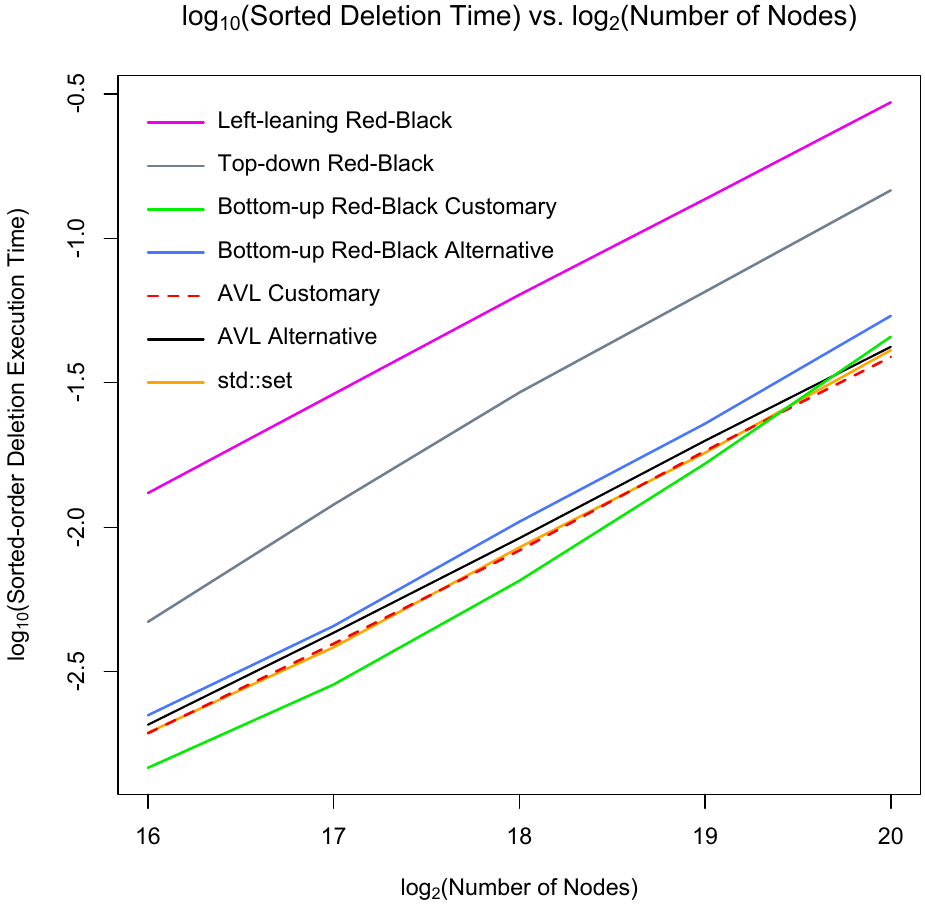}}
\caption{Sorted-order deletion execution times versus the number of nodes}
\label{fig:SortedDeletionTimes}
\end{figure}

Figure \ref{fig:SortedDeletionTimes} shows log-log plots of the Raptor Lake i7 CPU execution times for sorted-order deletion from each tree in the set for each type of tree and for \lstinline{std::set}, versus the number of nodes. The $y$-axis displacement between plots in log space, calculated from deletion times for $ 2^{18} $ nodes, reveals that the bottom-up red-black (green) tree's customary deletion algorithm is respectively $ 10^{0.10} = 1.27 $, $ 10^{0.15} = 1.40 $, $ 10^{0.20} = 1.60 $, $ 10^{0.99} = 9.76 $, and $ 10^{1.32} = 20.9 $ times faster than the AVL customary (red), AVL alternative (black), bottom-up red-black alternative (blue), top-down red-black (gray), and left-leaning red-black (magenta) trees' deletion algorithms; and also that the AVL tree's customary deletion algorithm (red) is $ 10^{0.04} = 1.10 $ times faster than its alternative deletion algorithm (black).

Hence, the bottom-up red-black tree outperforms the AVL tree, and the top-down and left-leaning red-black trees, for sorted-order deletion.

\newpage

The benchmark results for sorted-order insertion into and deletion from the AVL and red-black trees are summarized in Table \ref{table:SortedInsertionDeletionSpeed} that reports the relative speeds of the AVL and red-black trees for insertion and deletion of $ 2^{18} $ nodes. For insertion, the AVL tree is fastest. For deletion, the bottom-up red-black tree's customary deletion algorithm is fastest. The relative insertion speeds of the AVL and bottom-up red-black trees are consistent with the results of previously reported benchmarks \cite{Pfaff}.

% Create a narrower column type U to accommodate "Left-leaning" in the following table and
% then make a rightmost YYYY column to that table to suppress a rightmost vertical line
% that appears when the U column type is substituted for the default X column type.
% In fact, a rightmost YY column suffices but additional Y characters shrink the space
% between the rightmost character of the table and the surrounding tcolorbox.
\newcolumntype{U}{>{\hsize=0.60\hsize}X}

\begin{table}[htb]

\begin{tcolorbox}[tab2,tabularx={U||Y|Y|YYYY}]
Algorithm & Insertion & Customary Deletion & Alternative Deletion  \\\hline\hline
AVL & 1.00 & 1.27 &1.40  \\\hline
Bottom-up \newline Red-Black & 2.18 & 1.00 & 1.60 \\\hline
Top-down \newline Red-Black & 3.12 & 9.76  \\\hline
Left-leaning \newline Red-Black & 3.22 & 20.9  \\\hline
\end{tcolorbox}

\caption{\label{table:SortedInsertionDeletionSpeed}
Relative speeds for sorted-order insertion and deletion of $ 2^{18} $ nodes}

\end{table}

\section{Alternative Deletion}
\label{sec:AlternativeDeletion}

Table \ref{table:AVLBURBSortedRotationsVsN} shows the number of rotations required by customary and alternative deletion for the AVL and bottom-up red-black trees, and for $ 2^{16} $ through $ 2^{20} $ nodes.

% Create a narrower column type T to accommodate "Left-leaning" in the following table and
% then make a rightmost YYYY column to that table to prevent a line break in the Y and YYYY
% column headers, for some inexplicable reason.
\newcolumntype{T}{>{\hsize=0.95\hsize}X}

\begin{table}[htb]

\begin{tcolorbox}[tab2,tabularx={T||Y|Y|Y|Y|YYYY}]
Algorithm &  $2^{16}$ & $2^{17}$ & $2^{18}$ & $2^{19}$ & $2^{20}$  \\\hline\hline
AVL \newline Customary & 32,753 & 65,520 & 131,055 & 262,126 & 524,269  \\\hline
AVL \newline Alternative & 32,753 & 65,520 & 131,055 & 262,126 & 524,269  \\\hline\hline
Bottom-up \newline Red-Black \newline Customary & 32,754 & 65,521 & 131,055 & 262,127 & 524,270  \\\hline
Bottom-up \newline Red-Black \newline Alternative& 32,754 & 65,521 & 131,055 & 262,127 & 524,270  \\\hline
\end{tcolorbox}

\caption{\label{table:AVLBURBSortedRotationsVsN}
Rotations for AVL and bottom-up red-black customary and alternative deletion}

\end{table}

\newpage

This table reveals that, for sorted-order deletion from these two trees, both customary and alternative deletion require the same number of rotations. Hence, for sorted-order deletion, the attempts to select a preferred replacement node that are described in Sections \ref{sec:AVLDeletion} and \ref{sec:RedBlackDeletion} never select a preferred node and never avoid rebalancing; therefore, these attempts at selection merely prolong the deletion execution time without any beneficial reduction in the number of rotations.

Consequently, the log-log plots of Figure \ref{fig:SortedDeletionTimes} reveal the computational expense of attempted selection of a preferred replacement node. The $y$-axis displacement between plots in log space, calculated from deletion times for $ 2^{18} $ nodes, reveals that the customary deletion algorithms for the red-black (green) and AVL (red) trees are respectively $ 10^{0.20} = 1.60 $ and $ 10^{0.04} = 1.10 $ times faster than the alternative deletion algorithms for the bottom-up red-black (blue) and AVL (black) trees.

For the bottom-up red-black tree, selection of the preferred replacement node for a deleted two-child node requires (1) reading the \lstinline{size} fields of its two children and comparing their sizes, and (2) updating the \lstinline{size} field of any node involved in a rotation. The expense of these operations is revealed by the $y$-axis displacement between the bottom-up red-black tree customary (green) and alternative (blue) plots in Figure \ref{fig:RandomDeletionTimes}, and exceeds the benefit of avoiding rebalancing after random-order deletion.

For the AVL tree, selection of the preferred replacement node for a deleted two-child node requires reading that node's \lstinline{balance} field and testing its value. The expense of this operation is revealed by the coincident AVL tree customary (red) and alternative (black) plots in Figure \ref{fig:RandomDeletionTimes}, and approximately equals the benefit of avoiding rebalancing after random-order deletion.

\section{Cache Memory Effects}
\label{CacheMemory}

Table \ref{table:SortedInsertionDeletionRatio} reports the relative speeds of sorted-order and random-order insertion and customary deletion for the AVL and bottom-up red-black trees that comprise $ 2^{18} $ nodes, expressed as a ratio of those speeds.

% Create a narrower column type V to accommodate "Left-leaning" in the following table and
% then make a rightmost YYYY column to that table to prevent a line break in the Y and YYYY
% column headers, for some inexplicable reason.
\newcolumntype{V}{>{\hsize=0.60\hsize}X}

\begin{table}[htb]

\begin{tcolorbox}[tab2,tabularx={V||Y|YYYY}]
Algorithm & Insertion & Customary Deletion  \\\hline\hline
AVL & 6.1 & 9.6  \\\hline
Bottom-up \newline Red-Black & 2.7 & 9.6  \\\hline
\end{tcolorbox}

% For some unknown reason, using '$ 2^{18} $ or $ 2^{19} $' in the caption below causes the
% left border of the 'o' in 'or' to be partially clipped, so '\,\;' are included in math mode.
\caption{\label{table:SortedInsertionDeletionRatio}
Sorted-order/random-order speed ratio for insertion and deletion of $ 2^{18} $ nodes}

\end{table}

\newpage
 
Table \ref{table:InsertDeleteSortedRandomRotations} reports the number of rotations required by sorted-order and random-order insertion and customary deletion for $ 2^{18} $-node AVL and red-black trees.

% Create a narrower column type S to accommodate "Left-leaning" in the following table and
% then make a rightmost YYYY column to that table to prevent a line break in the Y and YYYY
% column headers, for some inexplicable reason.
\newcolumntype{S}{>{\hsize=0.60\hsize}X}

\begin{table}[htb]

\begin{tcolorbox}[tab2,tabularx={S||Y|YYYY}]
Algorithm & Insertion & Customary Deletion  \\\hline\hline
AVL \newline Sorted-order & 262,125 & 131,055  \\\hline
AVL \newline Random-order & 183,046 & 97,852  \\\hline\hline
Bottom-up \newline Red-Black \newline Sorted-order & 262,110 & 131,056  \\\hline
Bottom-up \newline Red-Black \newline Random-order & 152,705 & 99,403  \\\hline
\end{tcolorbox}

\caption{\label{table:InsertDeleteSortedRandomRotations}
Insertion and deletion rotations for $ 2^{18} $-node AVL and bottom-up red-black trees}

\end{table}

Tables \ref{table:SortedInsertionDeletionRatio} and \ref{table:InsertDeleteSortedRandomRotations} demonstrate that sorted-order insertion and deletion require more rotations than random-order insertion and deletion, but are faster. To understand this paradox, a third and a fourth series of benchmarks were performed, wherein the LLC cache load misses were measured via the Ubuntu \lstinline{perf stat -e LLC-load-misses} command. For the third series, the keys were inserted into and deleted from the AVL and red-black trees in random order. For the fourth series, the keys were inserted and deleted in increasing sorted order. Section \ref{sec:Methodology} indicates the numbers of keys and iterations for each execution in each series.

Table \ref{table:CacheMissRatios} summarizes the \lstinline{cpu_core/LLC-load-misses} for random-order insertion and deletion, divided by the \lstinline{cpu_core/LLC-load-misses} for sorted-order insertion and deletion, and for $ 2^{16} $ through $ 2^{20} $ nodes. This table reveals that random order causes between 10 and 87 times more \lstinline{LLC-load-misses} than sorted order, which explains the paradoxical improved performance of sorted-order relative to random-order insertion and deletion, despite the fact that sorted order requires more rotations.

% Create a wider column type Z to accommodate "Left-leaning" in the following table and
% then make a rightmost YYYY column to that table to suppress a rightmost vertical line
% that appears when the Z column type is substituted for the default X column type.
% In fact, a rightmost YY column suffices but additional Y characters shrink the space
% between the rightmost character of the table and the surrounding tcolorbox.
\newcolumntype{Z}{>{\hsize=1.1\hsize}X}

\begin{table}[htb]

\begin{tcolorbox}[tab2,tabularx={Z||Y|Y|Y|Y|YYYY}]
Algorithm &  $2^{16}$ & $2^{17}$ & $2^{18}$ & $2^{19}$ & $2^{20}$  \\\hline\hline
AVL & $24$ & $45$ & $43$ & $38$ & $74$ \\\hline
Bottom-up \newline Red-Black & $31$ & $10$ & $36$ & $40$ & $87$ \\\hline
\end{tcolorbox}

\caption{\label{table:CacheMissRatios}
Random-order/sorted-order \lstinline{LLC-load-misses} ratios for AVL and red-black trees}

\end{table}

A hypothesis to explain the disparity in \lstinline{LLC-load-misses} is as follows. For sorted-order insertion or deletion, two consecutive integer keys descend a tree via a similar path. Hence, the nodes that the first key causes to be loaded into cache during its descent of the tree are likely to remain resident in cache and are likely to be accessed by the second key. In contrast, for random-order insertion or deletion, it is unlikely that two consecutive keys descend the tree via a similar path. Hence, the nodes loaded into cache during the first key's descent of the tree are unlikely to be accessed by the second key. Thus, random-order likely causes more \lstinline{LLC-load-misses} than does sorted-order, and hence possibly demonstrates a cache memory effect.

\begin{figure}[h]
\centering
\centerline{\includegraphics*[trim = {0.97in, 3.47in, 1.36in, 1.51in}, clip, width=\textwidth]{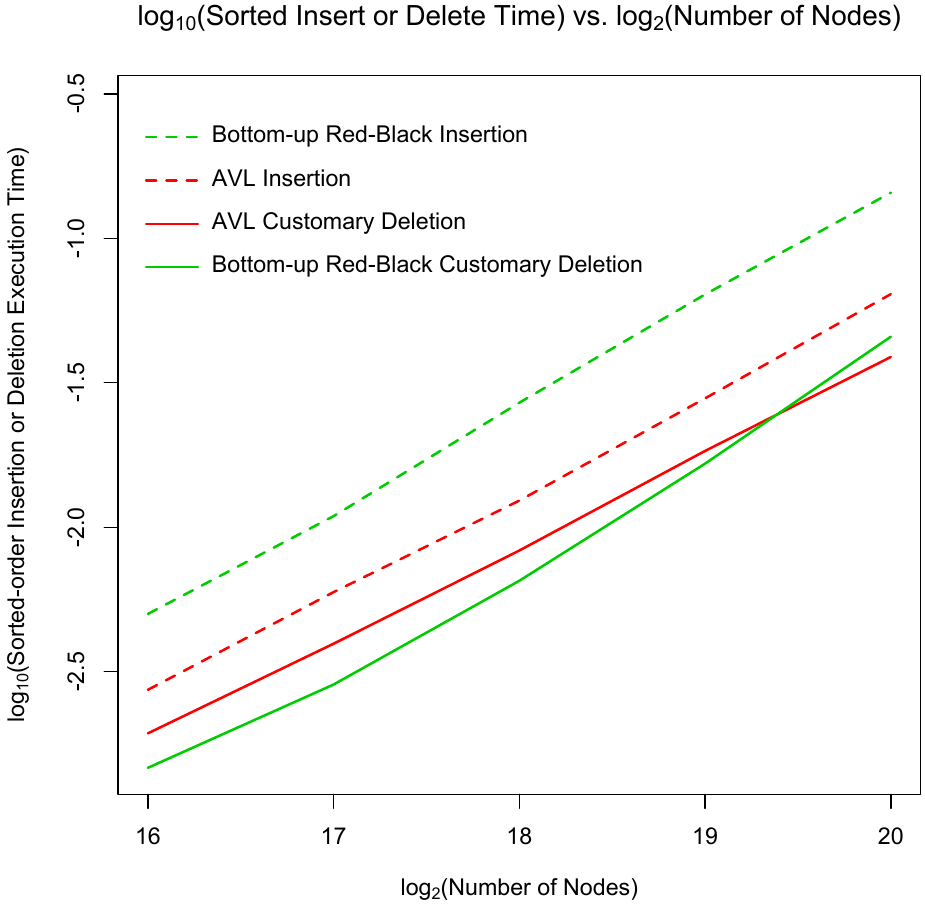}}
\caption{Sorted-order insertion and deletion execution times versus the number nodes}
\label{fig:SortedInsertionDeletionTimesVsN}
\end{figure}

Figure \ref{fig:SortedInsertionDeletionTimesVsN} suggests another possible cache memory effect. This figure shows log-log plots of the Raptor Lake i7 CPU execution times for sorted-order insertion into and customary deletion from the AVL and bottom-up red-black trees, versus the number of nodes.  

\newpage

This figure reveals that, for sorted order, the disparity between the insertion and deletion execution times is significantly greater for the bottom-up red-black tree than for the AVL tree. However, Table \ref{table:InsertDeleteSortedRandomRotations} reports that, for sorted order, the numbers of insertion and deletion rotations are nearly equal for these two trees. Hence, if the numbers of rotations explain the disparity between insertion and deletion execution times for the AVL tree, they likely cannot explain the disparity between insertion and deletion execution times for the bottom-up red-black tree.

Cache memory effects might contribute to the disparity between the insertion and deletion execution times for the bottom-up red-black tree. However, it is not possible to probe these cache memory effects via the Ubuntu \lstinline{perf stat -e LLC-load-misses} command, because that command measures the sum of LLC cache load misses for insertion and deletion. Instead, it would be necessary to measure the LLC cache load misses for insertion and deletion independently via, for example, the \lstinline{libpfm4} library \cite{Google}.

\section{Conclusions}
\label{Conclusions}

The alternative deletion algorithm proposed six decades ago \cite{Foster} reduces the number of rotations required after random-order deletion from the AVL and bottom-up red-black trees by 20 and 24 percent respectively. However, the computation required to implement this algorithm nullifies the benefit of fewer rotations.

The bottom-up red-black tree is faster than the AVL tree for insertion and deletion of randomly ordered keys. The AVL tree is faster than the bottom-up red-black tree for insertion but slower for deletion of consecutively ordered keys. The top-down red-black tree is faster than the bottom-up red-black tree for insertion but slower for deletion of randomly ordered keys, and slower for insertion and deletion of consecutively ordered keys. The left-leaning red-black tree is slower than the three other trees for insertion and deletion of randomly and consecutively ordered keys.

Random order and sorted order (i.e., consecutive order) bracket a variety of orders. Future work could investigate sequences in which sub-sequences of various lengths are randomly shuffled within an otherwise consecutively ordered sequence.

Future work could also investigate the disparity between insertion and customary deletion execution times for the AVL and bottom-up red-black trees, and probe whether that disparity is due to cache memory effects, by measuring LLC cache load misses independently for insertion and deletion.

\newpage

\section*{Supplemental Materials}

Included with this manuscript is a copy of Caxton Foster's 1965 article, ``A Study of AVL Trees," \cite{Foster} that was published as only a technical report internal to Goodyear Aerospace Corporation. Also included are C++ benchmark programs for the AVL tree and the bottom-up, top-down, and left-leaning red-black trees and \lstinline{std::set}.

Execution times for \lstinline{std::set} from the C++ Standard Template Library are measured via \lstinline{test_stdSet.cpp}.

The AVL tree implementation (\lstinline{avlTree.h} and \lstinline{test_avlTree.cpp}) was transcribed from Nicklaus Wirth's Pascal implementation of the AVL tree \cite{Wirth}. A bug in the \lstinline{del} procedure was fixed and that procedure was modified to create the \lstinline{eraseLeft} and \lstinline{eraseRight} functions that implement alternative deletion.

The bottom-up red-black tree implementation (\lstinline{burbTree.h} and \lstinline{test_burbTree.cpp}) was copied from Rao Ananda's C++ implementation of the bottom-up red-black tree \cite{Ananda}. The \lstinline{fixInsertRBTree} and \lstinline{fixDeleteRBTree} functions were renamed \lstinline{fixInsertion} and \lstinline{fixErasure} respectively and then modified to improve performance. Bugs and memory leaks were fixed in the \lstinline{fixDeleteRBTree} function. Iterative versions of the \lstinline{insert} and \lstinline{erase} functions were added that improve performance relative to the original, recursive versions of those functions. The \lstinline{erase} function was modified to implement alternative deletion.

The top-down red-black tree implementation (\lstinline{tdrbTree.h} and \lstinline{test_tdrbTree.cpp}) was transcribed from Cullen LaKemper's Java implementation of the top-down red-black tree \cite{LaKemper}. A bug was fixed in the \lstinline{removeStep2B2} method.

The left-leaning red-black tree implementation (\lstinline{llrbTree.h} and \lstinline{test_llrbTree.cpp}) was transcribed from Rene Argento's Java implementation of the left-leaning red-black tree \cite{Argento}. No bugs were discovered.

\section*{Acknowledgements}

I thank Paul McJones, Gene McDaniel, and John Robinson for helpful comments; Cullen LaKemper for discussions of the top-down red-black tree; Rene Argento for discussions of the left-leaning red-black tree; and Robert Tarjan for providing a copy of Caxton Foster's 1965 article, ``A Study of AVL Trees."

\section*{Author Contact Information}

\href{https://www.linkedin.com/in/russellabrown/}{https://www.linkedin.com/in/russellabrown/}

% explicitly declare a References section to compensate for removing "REFERENCES ... REFERENCES" in the jcgt.cls file.

\newpage

\section*{References}

\small
\bibliographystyle{jcgt}
\bibliography{paper}

\end{document}